\documentstyle[12pt,aasms4]{article}
\newcommand{\beq}{\begin{equation}}
\newcommand{\eeq}{\end{equation}}
\newcommand{\Mdot}{\dot{M}~}
\newcommand{\hii}{H~{\sc ii}~}
\newcommand{\kms}{\mbox{ km s$^{-1}$}~}

\newcommand{\gcm}{\mbox{ g cm$^{-3}$}~}

\newcommand{\Moy}{\mbox{M$_{\odot}$ yr$^{-1}$}~}

\begin{document}

\def\etal{{\it et~al.\ }}
\def\eg{{\it e.~g.\ }}
\def\ie{{\it i.~e.,\ }}

\title{Three Dimensional Magneto-Hydrodynamical Modeling of
Planetary Nebuale ~ II: The Formation of Bipolar and Elliptical Nebulae with 
Point-Symmetric Structures and Collimated Outflows}

\author{G. Garc\'{\i}a-Segura 
\\and\\
J. A. L\'opez}
\affil{Instituto de Astronom\'{\i}a-UNAM, Apdo Postal 877,
Ensenada, 22830 Baja California, Mexico;
ggs@bufadora.astrosen.unam.mx;
jal@bufadora.astrosen.unam.mx} 


\begin{abstract}
\noindent


This work presents three dimensional, magneto-hydrodynamical 
simulations of the formation and early evolution of a
subgroup of planetary nebulae that exhibit a variety of 
point-symmetric structures.
For bipolar nebulae, the formation of point-symmetric nebular shapes 
along the inner borders  of their opposing lobes and/or
collimated outflows or jets internal or external to their cavities,
is reproduced by considering a steady misalignment of the magnetic 
collimation axis with respect to the  symmetry axis of the bipolar wind
outflow, defined perpendicular to the equatorial density enhancement.
Elliptical planetary nebulae with ansae displaced from the
symmetry axis in point-symmetric fashion are reproduced through the same 
process by reducing the equatorial density enhancement.
This mechanism represents an alternative explanation to some cases
where morphological appearances give the impression  of the action of
a symmetric, rotating or precessing jet from the central source.
The computational survey reveals that jet formation is detected
only for dense enough winds with mass loss rates $\gtrsim 10^{-7}$ \Moy. 
For lower mass loss rates the jets tend to vanish leaving behind only 
ansae-like structures at the tips of the lobes, as observed in some cases.
The results are rather independent of the wind terminal velocity,
since magnetized bubbles behave adiabatically for low wind velocities
($\simeq$ $100 \kms$), which in the absence of a magnetic field would
behave as momentum driven.


\end{abstract}


\vfill\eject

\section{Introduction}

Numerical simulations that reproduce the shapes of planetary
nebulae (PNe) have helped to understand or give insights into the
basic physics acting behind the formation of the different
morphologies that are commonly observed in PNe. 
The two-wind model proposed by Kwok, Purton \& Fitzgerlad (1978)
has been applied and refined in a number of numerical simulations
based on hydrodynamic models (e.g. Khan \& West 1985; Mellema,
Eulderink \& Icke 1991; Frank \& Mellema 1994; Dwarkadas,
Chevalier \& Blondin 1996).
During the last decade, though, it has become evident that the
production of highly collimated, bipolar, supersonic outflows, play a 
major role in the evolution of a substantial number, if not all, of PNe 
(e.g. L\'opez 1997 for a review). 

The main nebular morphologies are usually well reproduced in
the hydrodynamic models, however, the production of jets and ansae
have met with certain difficulties; in particular, fast winds do not 
tend to converge into stable structures to form ansae and jets 
(e.g. Dwarkadas \& Balick 1998).  

On a somewhat different approach, Chevalier \& Luo (1994) have
explored the effects of a rotating star with a magnetized wind on
the formation of aspherical bubbles. Following this scheme, R\'o\.zyczka \&
Franco (1996) and Garc\'{\i}a-Segura et al. (1997a, 1999),
performed 2-D MHD simulations of this magnetized wind in
cylindrical and spherical calculations, respectively, showing
that magnetic tension may indeed be responsible for the generation 
of jets in PNe. Recently, Garc\'{\i}a-Segura (1997) (paper I) presented 
full 3-D MHD models of PNe where jets and ansae are convincingly
reproduced as the result of magnetic collimation of the post-AGB
wind.

A particularly intriguing case in PNe morphologies are
those that display point-symmetric structures. Point-symmetry has been  recognized in a wide variety of PNe (e.g. Guerrero, V\'azquez \& L\'opez 1998).
In some instances, where reliable kinematical information is
available, the presence of bipolar, rotating, episodic jets or
BRETS have been inferred to explain similar structures (c.f. L\'opez 1997).

Cliffe et al. (1995) performed 3-D numerical simulations
introducing in their computational mesh the presence of a BRET and
reproducing the main morphological features observed, in particular for
the case of Fleming 1 (L\'opez, Meaburn \& Palmer 1993). They also noted that some bipolar nebulae, such as Hb 5 present
distinct, clumpy, S-shaped intensity distributions in their opposing lobes, 
which led them to argue that these structures could also be originated by a
precessing jet. The numerical simulations presented in this paper have been
mainly motivated by this group of objects, though the results expand a wide
range of configurations, including the conditions for jet detection in
PNe, ansae formation in point-symmetric modes and `regular'
bipolar and elliptical PNe presenting some degree of asymmetry.

In \S~2 we discuss the initial considerations and the motivation for the
present paper. \S~3 describes the methodology. \S~4 presents results for
bipolar nebulae and \S~5 for ellipticals. \S~6 Discusses the results
and a summary is given in \S~7.

\section{Initial considerations} 

As an illustration of the main type of morphological structures
that are the focus of the present simulations, a mosaic of images of 
these type of nebulae is presented in Figure 1. The objects included in this
representative sample are, (1a) Hb 5, (1b) He 2-103, (1c) He 2-429 and (1d) NGC 2371. Hubble 5 is a bipolar nebula with distinct internal S-shaped intensity distribution. Recent HST images of this object obtained by Balick, Icke and Mellema show a great deal of detail in the secondary, tilted outflow. 
He 2-103 is an elliptical nebulae where the equatorial density enhancement is 
clearly not perpendicular to the elliptical shell; furthermore, faint nebular
extensions are discerned displaced from the main elliptical axis on both sides
of the nebula. He 2-429 (see Guerrero et al. 1998 and the IAC Catalog, Manchado
et al. 1996) is an elliptical nebula with twisted, point-symmetric extensions
protruding outside the main nebular shell. Finally, NGC 2371 has a dumb-bell
appearance with intensity maxima in the equator defining a minor axis which
is not orthogonal to the main nebular axis.

The main, standard, working hypothesis that sets the initial conditions of
the simulations is that the equatorial density enhancement (EDE) produced 
in the late AGB phase defines the main symmetry axis of the bipolar/elliptical
wind outflow, i.e. the EDE is orthogonal to the main bipolar/elliptical structure.
 
In paper I, in order to reproduce point-symmetric structures, a wide binary
nucleus was considered and the magnetic collimation axis (i.e. the
rotation axis of the primary star) was assumed to precess, with a period of
the same order of the kinematic age of the nebulae ($\sim 10^3$ yr).
 
In the present paper, instead of invoking an active rotation of the magnetically
collimated outflow, a misaligned configuration is produced by introducing a
steady tilt of the magnetic collimation axis with respect to the bipolar/elliptical 
wind outflow.  Hence, this is a particular case of paper I, in which the precession
period is nule or much larger than the kinematic age of the nebula. 

Thus, in order to achieve this steady misalignment, the rotating core of
the AGB, which will provide the magnetized fast wind, must have a rotation 
axis tilted with respect to the axis defined by the EDE. Our current limited
understanding of the detailed processes that produce the EDE at the 
AGB phase precludes a full physical justification for this assumption. 
Binary nuclei could represent the best way to attain this effect given the 
diversity of paths that they can take during their evolution through the 
AGB (e.g. Yungelson, Tutukov \& Livio 1993). For example, wide binary systems
may induce this misalignment since the ejected envelope's rotation axis of the
primary
star could be tidally altered by the secondary and the EDE formed in the orbital
plane, with a resulting tilt with respect to the rotation axis of the primary. 
Alternatively, systems which undergo a common envelope evolution, in which the 
rotation axis of the primary star is not orthogonal to the orbital plane could 
also produce the same effect. Close binary systems with substellar companions
that may lead to the formation of accretion disks (Reyes-Ruiz \& L\'opez 1999)
and mergers could also play a significant role here.  

However, beyond these arguments, which badly need additional
investigation, the best initial justification that has motivated at this 
time further exploring of this assumption  is the apparent asymmetry
observed in the equatorial regions of a substantial number of PNe
and the corresponding point-symmetric condensations and shapes (see Figure 1).

The numerical models do not consider an explicit physical form for the
source of the wind, just an outflowing magnetized wind  from 
a spherical radius of $r=0.005 $ pc after t=0. 
Regions below this limit lie beyond the
current resolution of the simulations.

The general idea behind these MHD models is sketched in Figure 2. 
Here, a rotating star 
produces the outflowing magnetized wind, in which the toroidal component of 
the magnetic field is dominant respect to the poloidal contribution at 
large distances.

\section{Three-dimensional MHD models}

The simulations have been performed using the magneto-hydrodynamic code 
ZEUS-3D (version 3.4), developed by M. L. Norman and the Laboratory for 
Computational Astrophysics. This is a finite-difference, fully explicit, 
Eulerian code descended from the code described in Stone \& Norman (1992). 
A method of characteristics is used to compute magnetic fields as described
in Clarke (1996).  ZEUS-3D does not include radiation transfer, but we have
implemented a simple approximation to derive the location of the ionization
front for arbitrary density distributions (see Bodenheimer \etal 1979; 
Franco \etal 1989, 1990, Garc\'{\i}a-Segura \& Franco 1996). This is done by assuming that ionization equilibrium
holds at all times, and that the gas is fully ionized inside the \hii region.

The location of the ionization front in any given direction $(\theta ,\phi )$
from the photoionizing source is given by 
$\int n^2(r,\theta,\phi) r^2 dr \approx F_{\star} / 4 \pi \alpha_B$, where
symbols have their usual meaning.
The models include the Raymond \& Smith (1977) cooling curve above $10^4$ K.
For temperatures below $10^4$ K, the shocked gas region is allowed to cool 
down with the radiative cooling curves given by Dalgarno \& McCray (1972) and
MacDonald \& Bailey (1981). Finally, the photoionized gas is always kept at $10^4$ K,
so no cooling curve is applied to the HII regions (unless, of course,
there is a shock inside the photoionized region). The minimum temperature
allowed in all models is set to $10^2$ K. The computations are done in 
Cartesian coordinates (x, y, z),  with outflowing outer boundary conditions. 

The models have grids of $100 \times 100 \times 200$ equidistant zones, 
with an extent of $0.1 \times 0.1 \times 0.2$ pc in x, y and z respectively.  
The wind, whose origin is centered at the center of the computational
volume, is set based on the rotating wind solutions given by 
Bjorkman \& Cassinelli  (1993), Ignace et al. (1996, 1998)
and the approach to those equation given by Garc\'{\i}a-Segura et al.(1999).
These equations permit the introduction of the wind self-consistently. 
For the reasons given in the preceding section, in the present calculations we
do not address a particular functional shapes for the EDE, but simply state its
presence independently of its origin.

The  magnetic field of an outflowing wind from a rotating star
can be described by two components, $B_{\phi}$ and $B_{\rm r}$ 
(Chevalier  \& Luo 1994; R\'o\.zyczka \& Franco 1996; paper I;
Garc\'{\i}a-Segura et al. 1999).
The radial field component  can be neglected since $B_{\rm r}
\sim r^{-2}$ while the toroidal component $B_{\phi} \sim r^{-1}$, and the
field configuration obeys $ \nabla \cdot B = 0 $ .
In all the simulations,  the toroidal field is set always
perpendicular to the z axis, therefore   
only $B_{{\phi}_{\rm x}}$ and $B_{{\phi}_{\rm y}}$
are introduced in the source wind sphere, with $B_{{\phi}_{\rm z}}=0 $. 

As in paper I, we will use two useful dimensionless parameters
containing the basic information of the models.
The first one $\Omega$ is the velocity ratio of the stellar rotation to the
critical rotation, and the second parameter $\sigma$ is the ratio of 
the magnetic field energy density to the  kinetic energy density in the 
fast wind.
The wind compression field (``WCField'') found by Ignace et al.(1998) 
is also introduced.  However, since $\Omega$ is less than 10\% for the 
fast wind in all the calculations, this correction is of little relevance 
in these cases.
 
The AGB wind is set with a $v_{\infty}=10 \kms$ and a $\Mdot = 10^{-6}
\Moy $ as a standard value in all the models. The AGB wind
is set unmagnetized in all models, because the estimated values for
$\sigma$ are very low in this regime (see paper I).
Since we are interested in the early phases of the PN evolution,
the velocity of the fast wind is set to $100 \kms$. 
This value should reasonably match the conditions at the departure from
the AGB phase. 

A more realistic approach to this problem  would need to include 
the evolving wind conditions with gradual transitions of the wind from
$v_{\infty} \sim 10 $ \kms up to $ \sim 1000 $ \kms (or higher), 
and mass-loss rates varying from $\Mdot = 10^{-6} $ \Moy  (or higher)  
down to $ 10^{-9} $ (or lower).
In this case time dependent effects invoke additional behaviors
on the dynamics (e.g. Dwarkadas \& Balick 1998). They are out of the
scope of the present paper and are deferred to a future work.

The  parameters of the runs are summarized in Table 1. These include, 
an unmagnetized reference case (model A0); bipolars with  different
inclinations of the bipolar wind outflow symmetry axis with respect to the 
magnetic collimation axis and two different mass loss rates (models
B and C, respectively) and ellipticals with a single constant tilt and three
different mas loss rates (models E)

\section{Bipolar Nebulae}

Bipolar nebulae are mainly characterized by an equatorial waist with
a substantial pole to equator density contrast. In order to achieve
such a density contrast, we have used a value of   $\Omega=0.98$ in
the set up, which is obtained with a rotational velocity of 6.929 \kms.

\subsection{The unmagnetized bipolar case}

As a starting point, the response of a spherically symmetric, unmagnetized 
wind starting from a spherical shape is shown in model A0 (Fig. 3).
This unmagnetized case shows how the wind behaves in such a medium, i.e., 
the wind is impeded to expand in the equator due to the EDE, but it
easily expands along the polar direction. This would be the classical hydrodynamic case that has been thoroughly studied in the past 
(c.f. see Frank 1999 for a review on bipolar outflows).
Because of the radiative conditions of the reverse or terminal shock of such a
wind in this set up 
( $t_{\rm post-shock} \sim 10^5$ K for $v_{\rm shock}= 100 \kms$), 
the bipolar bubble is momentum driven.
This particular dynamics generates ram-ram pressure instabilities 
(Garc\'{\i}a-Segura et al. 1996; Garc\'{\i}a-Segura et al. 1997b), however,
given the low resolution of the grid, these instabilities are not 
reproduced in the present study, and only some indications of such a 
behavior is present 
as the small corrugation that appears in the lobes of model A0.

Model A0 is important as a reference model since it shows that fast
unmagnetized  winds, starting from spherical conditions and expanding within 
an EDE  cannot converge to form jets or ansae.
In more adiabatic conditions, the lobes become two tangential spheres 
(Garc\'{\i}a-Segura et al. 1999). It is precisely this behavior that rules
under the presence of an EDE, i.e. the bipolar lobes will grow always perpendicular 
to the EDE, and is the EDE the responsible for the equatorial
waist observed in bipolar nebulae. 

\subsection{Magnetized Bipolar Nebulae}

Keeping in mind the behavior of the reference model A0, a magnetized wind is
now introduced to visualize the strong change that effects the dynamics.
Model B0 (Fig. 3) only differs from model A0 in the magnetized wind, with
a $\sigma=0.01$. But the result is remarkable, a pair of jets form at
both sides in the lobes. The explanation of this behavior is fully described
in R\'o\.zyczka \& Franco (1996), and it results mainly from the 
action of the hoop stress caused by the toroidal magnetic field.
The rest of the bubble behaves now adiabatically because the magnetic
pressure does not allow the gas to pile up towards the inner edges 
of the lobes. This is also the main reason why model B0 expands much faster
than model A0, since energy conserving bubbles in power laws of the type
$r^{-2}$ have different dynamics than momentum conserving bubbles 
(Garc\'{\i}a-Segura \& Mac Low 1995 and references therein).

Although the hoop stress is always present in this type of flows,
not always the conditions are in favor to form observable jets. Such a case is
computed in model C0 (Fig. 3), in which the mass-loss rate of the
fast wind has been decreased by an order of magnitude. In such conditions,
the fast wind piles up at the poles, but the radiative cooling, proportional 
to $n^2$, is not sufficiently strong to allow the formation of visible jets,
and ansae-like structures are formed instead.
This difference is more evident in the synthetic pictures shown in 
figure 7 for models Bs and Cs, which are discussed further below.

As mentioned before, the symmetry axis of the bipolar wind outflow is 
considered orthogonal to the EDE. Models B0 and C0 are computed with the 
magnetic collimation axis also orthogonal to the EDE. The following models 
break this condition by introducing
different tilts between these axes to explore the resultant 
effects, which are shown to lead to point-symmetric shapes.

The tilt is introduced at the EDE for simplicity. Thus, the magnetic
collimation axis remains parallel with the z axis, and the same
set up described above is used for the magnetized wind.

In order to have a general picture of the effect caused by the tilt,
we have computed several models in which the tilt angle varies from low
values, $ 5^{\circ}$ (models B5 and C5), middle values, 
$15^{\circ}$ (models B15 and C15) and large values, $45^{\circ}$
(models B45 and C45).  The evolution of the two last models is shown 
in figures 4 and 5, respectively.

The dynamics of the bubbles in such conditions are quite complex
and the limited resolution of the models hinder a full, quantitative
description. Nevertheless, the effect of the tilt can be summarize 
qualitatively as follows.

At the beginning of every simulation, the spherically symmetric shocked flow 
is driven by the magnetic pressure gradient after crossing the reverse or
terminal shock and  is launched in the z direction on both sides of the wind 
source region. Since the  magnetic field is toroidal, the magnetic pressure
is nule towards the z axis in the free expanding wind (but not in the 
shocked region), while it is maximum in the equatorial plane, consequently
building a  large pressure gradient. Gradually, the flow notices the effect of 
the EDE and is deflected perpendicular to it. The magnetically pressurized 
bubble is distorted, and the flow, which now follows the tendency of the bipolar
cavity, pushes and deflects the column of gas in which the hoop stress is
larger. When this column is dense enough, the formed jets are deflected
an acquire the appearance of  precessing jets, which eventually will
collide with the bipolar cavity (see e,g, model B45). 
Thus, there are two competing effects, one is the tendency of
the gas to flow perpendicular to the EDE, and the other is the tendency
of the gas to flow parallel to the z axis as imposed by the magnetic
pressure and the hoop stress. When the tilt is nule, i.e., when
both effects do not compete with each other, the bubble acquires the largest
expansion velocity at the polar regions.

\section{Magnetized Elliptical Nebulae} 

Elliptical nebulae are characterized by a moderate pole to equator density 
contrast. We have adopted a value of  $\Omega=0.80$  
for the AGB wind, which is obtained with a rotational velocity of 5.65 \kms.
The AGB wind is set with a $v_{\infty}=10 \kms$ and a $\Mdot = 10^{-6} \Moy $ 
as a standard value for models E1 and E2, while $\Mdot = 10^{-8}\Moy $
is used for E3 and E4. The mass loss rate of the fast wind is varied from 
values of $\Mdot = 10^{-7} \Moy $ (E1) to  $\Mdot = 10^{-9} \Moy $
(E3 and E4). A tilt of $15^{\circ}$ is used in this case for all models for 
simplicity.

An example of the model evolution is shown in Fig. 6 for model E1.
The evolution is similar to the models described above for bipolar nebulae,
with the difference that the bubble grows further at the equator
since the EDE is not as restrictive as before. Figure 7 (right panels)
shows the apparent differences between models E1, E2, E3 and E4, dictated
by varying the mass loss rates of the fast wind from relatively large to 
lower values.
The polar, piled-up gas in model E1 ($\Mdot = 10^{-7} \Moy$) is able to cool
down efficiently in a dynamical time scale comparable to the computed time
and the axial jets become readily apparent. In model E2  
($\Mdot = 10^{-8} \Moy$) the cooling is less efficient but still present, 
giving rise to the formation of ansae-type structures. Models E3 and E4 
($\Mdot = 10^{-9} \Moy$) do not show any apparent signs of piled-up gas
(see Figure 7). 
It should be noted, however, that given the limited numerical resolution
of these models,
the cooling regions are not resolved and the mass loss rates should be taken
as estimative values. However, since these models have energy conserving
dynamics (as imposed by the magnetic pressure) rather than momentum conserving,
these estimations for the mass loss rates are quite independent of the terminal
wind velocities. In the absence of a magnetic field, winds of the order of 
$100 \kms$ have strongly radiative terminal or reverse shocks 
(see for example Frank et al. 1996), and the radiative conditions are strongly 
dependent on the wind velocity and not only of the mass loss rate.

\section{Analysis of the models}

In order to test the numerical results of the models, it would be
convenient to run convergence tests at lower and higher resolution. The latter
becomes prohibitively demanding for our computational resources.
An experiment at half the resolution, which gives a qualitative indication
on the reliability of the model is presented in Figure 6b.
This figure shows a comparison for model E1 with two different
resolutions.  Althought, the individual details in both simulations in that
figure differ by obvious reasons, the main features, such as the shape 
of the bubble
and the formation and extension of the jets remain rather similar.
The densities are larger in the high resolution run, indicating 
that in order to make quantitative estimates of physical parameters,
one needs twice the resolution 
at least to be sure that the dynamics has been resolved, i.e., to find a 
convergence. 
For example, what is the post-shock cooling length
and how does this compare with the resolution of
the simulations ?

In order to resolve a cooling distance, one needs basically to get
$ d_{\rm cool}   >  x_{\rm max} / n_{\rm x}  $ , where $x_{\rm max}$ is the
physical size of the grid, and $n_{\rm x}  $ the number of zones  (i.e.  
$d_{\rm cool}$ bigger than one zone).
However, several zones are needed to resolve detailed cooling processes 
(e.g., any sort of microphysics).
A rough estimate of
the cooling timescales, can be obtained from equilibrium line cooling.
In this case,  the cooling time of a parcel of shocked gas is given by
$t_{\rm cool} = 3^{3/2} \,\,v_{\rm s}^3/(2^8 \,\,A \,\,\rho_0) \simeq 0.3
\,\,v_8^3 \,\,n_6^{-1}$ yr (Franco \etal 1994), where $v_8 = v_{\rm s}/10^8$
cm s$^{-1}$ is the shock velocity, $\rho_0$ is the pre-shock mass gas density
and $n_6$ is this density in units of $10^6$ cm$^{-3}$, $A=3.9\times 10^{32}
\,\,q_{\rm p}^{-5/2}$ cm$^6$ s$^{-4}$ g$^{-1}$, and $q_{\rm p}$ is the mass
per particle in units of $m_{\rm H}$ (\ie $q_{\rm p}=1.4/2.3$ for a fully
ionized gas with cosmic abundances).
For typical numbers in our simulations ($v_{\rm s} \sim 100 \kms$, 
$ n \sim 10 $ cm$^{-3}$) we obtain $t_{\rm cool} \sim 30 $ yr, which give
us $ d_{\rm cool} \sim 2 \times 10^{15} $ cm, just slightly smaller than the
zone zise of $3\times 10^{15} $ cm. 
One should also consider that in MHD models conditions are such that 
cooling distances are more favourable resolved perpendicular to the field
lines given the tendency to adiabatic behaviors in this case.

Since the MHD simulations behave similar to the hydro solutions along the Z axis
(parallel to the field lines), except for the extra pressure (hoop stress)
that the magnetic field produces towards the rotation axis of the central star,
it is possible to use the relations of conical converging flows
(Cant\'o, Tenorio-Tagle \& R\'o\.zyczka 1988; or the specific solutions for
PPNe by Frank, Balick \& Livio 1996), as an approximation for the
parameters in the jets. 
This is in fact a conical converging, magnetized flow problem.
Thus, for example, the density in the jets can be estimated by following 
formula (23) in Cant\'o, Tenorio-Tagle \& R\'o\.zyczka (1988), where
$ \rho_{\rm j} \simeq 6 \times 10^3 sin^4 \theta
\left[ v_{0} / 100 \kms \right]^4 \rho_{0} $ . For the case of
our E1 model, one gets 
(for $\theta \sim 50^{\circ}$) $ \rho_{\rm j} \sim 1  \times 10^3 \rho_{0} $.
On the other hand, the computed models give 
$ \rho_{\rm j} \sim 1 \times 10^1 \rho_{0} $ for the low resolution run
and $ \rho_{\rm j} \sim 1 \times 10^2 \rho_{0} $ for the higher resolution one.
Since the resolution is not big enough (as discused above), and in both cases,
 $\rho_{0} \sim 1 \times 10^{-23} $ \gcm, it is expected to find  a jump
of three orders of magnitude on the density ratio at twice the resolution, which
would match the analytical value.

\section{Discussion}

The computational survey reveals that the formation of jets and ansae, or none,
can be achieved by varying the mass-loss rate within a common framework of magnetized winds. In principle this would allow to have an estimate of the
mass-loss rates involved in the formation of the different types of collimated
outflows found in PNe. In particular, it is found that for the formation and
detection of jets, the mass-loss rates must have values 
$\Mdot \gtrsim 10^{-7}~ \Moy $. In the case of ansae a 
$\Mdot \sim 10^{-8}~ \Moy $ is required. Below these values, i.e. for typical
cases of $\Mdot \sim 10^{-9}~ \Moy$ or lower, no collimated features are
formed or revealed by the models. These numbers, however, must be taken with
caution given the  low resolution used in these 3-D calculations.
Nevertheless, there are additional studies that indicate that jets in PNe
must be formed during a brief lapse of high mass-loss rate in the pre-planetary
nebula stage. For example, Reyes-Ruiz \& L\'opez (1999) have concluded that
when an accretion disk is involved in the formation of collimated outflows in PNe, the necessary high accretion rates, $\Mdot \sim 10^{-7}$, 
needed to launch an axis-symmetric wind (as in the case of young stellar
objects, e.g. Calvet 1998) are attained only for a few thousand years during
the late AGB or pre-planetary phase.
 
Figure 7 shows different models in which jets have formed. The projected
shapes in the tilted cases give the impression of a  varying 
direction in the outflow, closely resembling the point-symmetric structures often observed in a number of  PNe. However here, no active precession or rotation in the source of the outflows is involved. These shapes result by 
the steady misalignment of the magnetized wind's axis with respect to the 
axis of the  bipolar wind outflow. This configuration can lead to a hydrodynamical deflection of the magnetized wind on the bipolar cavity. 
This is a novel feature revealed by the simulations and which can be tested in
the future by detailed kinematical and spectropolarimetric studies. For example,
in this case the deflected gas is expected to show a kinematic behavior that deviates from the rest of the nebular expansion and line emission in the
magnetically collimated regions to show polarization effects. 
Hydrodynamical deflection can only account for S-shaped morphologies, helicoidal
type structures would require some sort of rotation or wobbling from the wind source.

Although the PNe that were originally classified as point-symmetric comprise
a very reduced group, the presence of point-symmetric structures in PNe
has been shown now to span a much wider sample (Guerrero, V\'azquez \& L\'opez
1998). For example, signatures of point-symmetry can be appreciated in
nearly 20\% of the sample contained in the IAC morphological catalog 
(Manchado et al. 1996). Furthermore, practically every PN that has been
recently observed at high spatial resolution with the Hubble Space Telescope
shows some degree of point-symmetry, indicating that the mechanisms that
produce it should be rather common. The representative models shown in 
Figures 2 - 7 reveal the variety of point-symmetric configurations that 
can be formed under the assumptions and conditions of the present study. 
The inclusion of binary nuclei in this analysis can only but strengthen the
conditions for the development of point-symmetry.

\section{Summary} 

We have made a computational survey of 3-D MHD simulations of young
planetary nebulae in which are introduced different steady misalignments
of the magnetic collimation axis with respect to the symmetry axis of the bipolar/elliptical wind outflow. The latter defined as perpendicular to the
equatorial density enhancement. This steady tilt can also be interpreted
as the result of a very large precession period. The simulations show that
in these cases a hydrodynamical deflection of the magnetized collimated wind 
on the bipolar/elliptical cavity can produce morphologies which may resemble
the presence of precessing or rotating sources. In this way, the resulting morphologies are able to reproduce a considerable number of PNe which present some degree of point-symmetry in their structures.
The survey also suggests that the formation and detection of jets and ansae, 
or none, strongly depends on the mass-loss rate of the driving,
magnetized wind. In particular, the presence of jets is detected in the 
models only for mass-loss rates $\Mdot \gtrsim 10^{-7}~ \Moy $; below these
rates the jets tend to vanish in the models, keeping for $\Mdot \sim 10^{-8}~ \Moy $ only ansae type structures.

\bigskip
\bigskip

Acknowledgements

GGS is  grateful to J. Franco, N. Langer, M.-M. Mac Low, A. Manchado and 
M. Peimbert for useful discussions. 
We thank our anonymous referee for his comments.
We also thank Michael L. Norman and the Laboratory for Computational 
Astrophysics for the use of ZEUS-3D. The computations
were performed at the Instituto de Astronom\'{\i}a-UNAM.
This work has been supported by grants from DGAPA-UNAM (IN 114199) 
and  CONACyT (32214-E).

\begin{center} List of References \end{center}
\begin{description}

\item Bjorkman, J. E., \& Cassinelli, J. P. 1993, ApJ, 409, 429
\item Bodenheimer, P., Tenorio-Tagle, G. \& Yorke, H. W. 1979,
ApJ, 233, 85
\item Calvet, N. 1998, in Accretion Processes in Astrophysical Systems:
Some like it hot!, ed. S.S. Holt \& T.R. Kallman, (AIP press: New York), p. 495
\item Cant\'o, J., Tenorio-Tagle, G., \& R\'o\.zyczka, M. 1988, A\&A, 192, 287
\item Chevalier, R. A., \& Luo, D. 1994, ApJ, 421, 225
\item Cliffe, J. A., Frank, A., \& Jones, T.W. 1996, M.N.R.A.S., 282, 1114 
\item Clarke, D. A. 1996, ApJ, 457, 291
\item Dalgarno, A. \& McCray, R. A. 1972, ARAA, 10, 375
\item Dwarkadas, V. V., Chevalier, R. A. \& Blondin, J.M. 1996, ApJ, 457, 773
\item Dwarkadas, V. V. \& Balick, B. 1998, ApJ, 497, 267
\item Franco, J., Miller, W. W., Arthur, S. J., Tenorio-Tagle, G. \&
Terlevich, R. 1994, ApJ, 435, 805
\item Franco, J., Tenorio-Tagle, G. \& Bodenheimer, P. 1989, RMxAA, 18, 65
\item Franco, J., Tenorio-Tagle, G. \& Bodenheimer, P. 1990, ApJ, 349, 126
\item Frank, A. 1999, New AR, 43, 31
\item Frank, A. \& Mellema, G. 1994, A\&A, 289, 937
\item Frank, A., Balick, B., \& Livio, M. 1996, ApJ, 471, L53
\item Garc\'{\i}a-Segura, G. 1997, ApJL, 489, L189 (paper I)
\item Garc\'{\i}a-Segura, G., \& Franco, J. 1996, ApJ, 469, 171
\item Garc\'{\i}a-Segura, G., Langer, N., and Mac Low, M.-M. 1997b,
''LBV Outbursts: The Effects of Rotation", in A.S.P. Conference Series, 
{\em Luminous Blue Variables: Massive Stars in Transition}, ed. A. Nota and 
H. Lammers, Vol 120, 332
\item Garc\'{\i}a-Segura, G., Langer, N., R\'o\.zyczka, M., \&
Franco, J. 1999, ApJ, 517, 767
\item Garc\'{\i}a-Segura, G., Langer, N., R\'o\.zyczka, M., Mac
Low, M.-M., \& Franco, J. 1997a, in IAU Symp. No 180 ``Planetary Nebulae'',
eds. H.J. Habing  \&  H.J.L.M. Lamers, (Kluwer Academic Publishers, Dordrecht), 226
\item Garc\'{\i}a-Segura, G., Mac Low, M.-M., \& Langer, N. 1996, A\&A, 305, 229
\item Garc\'{\i}a-Segura, G., \&  Mac Low, M.-M. 1995,  ApJ, 455, 145 
\item Guerrero, M.A., V\'azquez, R., \& L\'opez, J.A. 1998, ApJ, 117, 967
\item Ignace, R., Cassinelli, C. P., \& Bjorkman, J. E. 1996, ApJ, 459, 671
\item Ignace, R., Cassinelli, C. P., \& Bjorkman, J. E. 1998, ApJ, 505, 910 
\item Khan, F.D. \& West, K.  1985, MNRAS, 212, 837 
\item Kwok, S., Purton, C. R. \& Fitzgerald, P. M. 1978, ApJ 219, L125
\item L\'opez, J. A. 1997,  in IAU Symp. No 180 ``Planetary
Nebulae'', eds. H.J. Habing  \&  H.J.L.M. Lamers, (Kluwer Academic Publishers, Dordrecht), 
197
\item L\'opez, J.A., Meaburn, J. \& Palmer, J.  1993, ApJ, 415, L135
\item MacDonald, J. \& Bailey, M. E. 1981, MNRAS, 197, 995
\item Manchado, A., Guerrero, M., Stanghellini, L., \&
Serra-Ricart, M. 1996, 
``The IAC Morphological Catalog of Northern Galactic Planetary
Nebulae'',
ed. Instituto de Astrof\'{\i}sica de Canarias
\item Mellema, G., Eulderink, F. \& Icke, V. 1991, A\&A, 252, 718
\item Raymond, J. C. \& Smith, B. W. 1977, ApJS, 35, 419
\item Reyes-Ruiz \& L\'opez 1998, ApJ, 524, 952
\item R\'o\.zyczka, M. \& Franco, J. 1996, ApJL, 469, L127
\item Stone, J. M. \& Norman, M. L. 1992, ApJS, 80, 753
\item Yungelson, L. R., Tutukov, A. V., \& Livio, M. 1993, ApJ, 418, 794

\end{description}

\newpage 
\begin{center} {\bf Figure Captions}
\end{center}
\begin{description}

\item Figure 1.  A representative sample of the type of nebulae that are
the focus of this study. 1(a) Hb 5; 1(b) He 2-103; 1(c) He 2-429; 1(d) NGC 2371.
See models B45, E3, E1 and E4 in Figure 7, respectively for comparison.

\item Figure 2.  Schematic view of the models. The magnetic collimation
axis (dashed line) is tilted with respect to the symmetry axis of
the bipolar wind outflow, which is perpendicular to the equatorial density
enhancement.

\item Figure 3. Evolution of the logarithm density distribution (g cm$^{-3}$) 
of a two-dimensional slice in the  x-z plane which crosses through the middle
of the computational volume. Top panels correspond to the unmagnetized case, model A0 ($\Mdot=10^{-7}~ \Moy$) for t = 1250 and 2500 yr. Middle panels correspond to the magnetized case, model B0 ($\sigma = 0.01$ and 
$\Mdot=10^{-7} \Moy$) for t = 900 and 1800 yr. Bottom panels, magnetized case, model C0 ($\sigma = 0.01$ and $\Mdot=10^{-8}~ \Moy$) for t = 1250 and 2500 yr. In all these cases the symmetry axis of the bipolar wind outflow and the magnetic collimation axis are coincident.

\item Figure 4. Evolution of the logarithm density distribution (g cm$^{-3}$) 
of a two-dimensional slice in the x-z plane which crosses through the middle 
of the computational volume for model B45 ($\sigma = 0.01$ and $\Mdot=10^{-7} ~\Moy$). Here the symmetry axis of the bipolar wind outflow is tilted by $45^{\circ}$ with respect to the magnetic collimation axis (t = 350, 700, 1000, 1350, 1650, 2000 yr).

\item Figure 5. Same as Figure 4 but for model C45 where $\Mdot=10^{-8} ~\Moy$ (t = 400, 850, 1250, 1650, 2100, 2500 yr).

\item Figure 6.  Same as Figure 4 for model E1 ($\sigma = 0.01$ and 
$\Mdot=10^{-7} ~\Moy$). Here the symmetry axis of the bipolar wind outflow is tilted by $15^{\circ}$ with respect to the magnetic collimation axis
(t = 250, 550, 800, 1050, 1350, 1600 yr).

\item Figure 6b. Two-dimensional slices of Model E1 for $50 \times 50 
\times 100$ (top) and for $100 \times 100 \times 200$ (bottom) 
equidistant zones in x, y and z respectively.

\item Figure 7. Two-color, composite, synthetic images of the
models where red represents the shock-dominated (non-photoionized)
material and
green the photoionized gas. The line of sight corresponds to 
the y axis, being the x-z plane the plane of the sky.
Bipolar models in the left panels show the magnetized case
with no tilt between the symmetry axis of the bipolar wind outflow and
the magnetic collimation axis (model B0) and the magnetized cases 
for three different tilts (B5, B15 and B45). In these models 
$\Mdot = 10^{-7} ~\Moy$ and $\sigma = 0.01$.
The central panels correspond to the bipolar models, as before, but 
with $\Mdot = 10^{-8} ~\Moy$. Elliptical models are shown in the right panels.
A tilt of $15^{\circ}$ is present in these four models. E1 
($\Mdot = 10^{-7} ~\Moy$, $\sigma = 0.01$), E2 ($\Mdot = 10^{-8} ~\Moy$,
$\sigma = 0.01$), E3 ($\Mdot = 10^{-9} ~\Moy$, $\sigma = 0.01$) and
E4 ($\Mdot = 10^{-9} ~\Moy$, $\sigma = 0.99$). Each run in the figure
corresponds to: B0 and B5, 1800 yr; B15 and B45, 2000 yr; 
C0, C5, C15 and C45, 2500 yr; E1, 1600 yr; E2, 1550 yr;
E3, 2000 yr and E4, 350 yr.

\end{description}

\newpage
 
\begin{table}  
\centering
\caption[junk]{Wind Properties}
\begin{tabular}{lcccccc}
\hline
\hline
Model & $\Mdot$ & $v_{\rm \infty}$ & $B_{\star}$&  $\sigma$ &
$\log F_{\star}$ & $\alpha$ \\
$\,$  & $\Moy$  &      $\kms$     &   G   &  $\,$   & s$^{-1}$ & 
deg \\  
\hline
$\,$  & $\,$ & Unmagnetized & Bipolar &  Model & $\,$ &  $\,$ \\
\hline
A0    &$1 \times 10^{-7}$ & 100.  &   0       &  0.00 &    44  & 0   \\
\hline
$\,$  & $\,$ & Magnetized & Bipolar & Models & $\,$ & $\,$\\
\hline
B0    &$1 \times 10^{-7}$ & 100.  &   16      &  0.01 &    44  & 0  \\
B5    &$1 \times 10^{-7}$ & 100.  &   16      &  0.01 &    44  & 5  \\ 
B15   &$1 \times 10^{-7}$ & 100.  &   16      &  0.01 &    44  & 15 \\ 
B45   &$1 \times 10^{-7}$ & 100.  &   16      &  0.01 &    44  & 45 \\ 
C0    &$1 \times 10^{-8}$ & 100.  &    5      &  0.01 &    43  & 0  \\ 
C5    &$1 \times 10^{-8}$ & 100.  &    5      &  0.01 &    43  & 5  \\ 
C15   &$1 \times 10^{-8}$ & 100.  &    5      &  0.01 &    43  & 15 \\ 
C45   &$1 \times 10^{-8}$ & 100.  &    5      &  0.01 &    43  & 45  \\
\hline
 $\,$ & $\,$  & Magnetized  & Elliptical  & Models  & $\,$ & $\,$\\
\hline
E1    &$1 \times 10^{-7}$ & 100.  &   16      &  0.01 &    44  &  15 \\
E2    &$1 \times 10^{-8}$ & 100.  &    5      &  0.01 &    43  &  15 \\
E3    &$1 \times 10^{-9}$ & 100.  &    2      &  0.01 &    42  &  15 \\
E4    &$1 \times 10^{-9}$ & 300.  &  104      &  0.99 &    42  &  15 \\
\hline
\end{tabular}
\end{table}

\end{document}